\documentclass[twocolumn,pre,showpacs,preprintnumbers,amsmath,amssymb,floatfix,superscriptaddress]{revtex4-1}
\usepackage{graphicx}
\usepackage{amsmath}
\usepackage{amssymb}
\usepackage{color}

\begin{document}

\title{Depletion interaction effects on the tunneling conductivity of nanorod suspensions}

\author{B. Nigro}\affiliation{LPM, Ecole Polytechnique F\'ed\'erale de
Lausanne, Station 17, CP-1015 Lausanne, Switzerland}
\author{C. Grimaldi}\affiliation{Laboratory of Physics of Complex Matter, 
Ecole Polytechnique F\'ed\'erale de Lausanne, Station 3, CP-1015 Lausanne, Switzerland}
\author{M. A. Miller}\affiliation{Department of Chemistry, Durham University, South Road,
Durham DH1 3LE, United Kingdom}
\author{P. Ryser}\affiliation{LPM, Ecole Polytechnique F\'ed\'erale de
Lausanne, Station 17, CP-1015 Lausanne, Switzerland}
\author{T. Schilling}\affiliation{Universit\'e du Luxembourg, 162 A, avenue de la
Fa\"iencerie, L-1511 Luxembourg}

\begin{abstract}
We study by simulation and theory how the addition of insulating spherical particles affects
the conductivity of fluids of conducting rods, modeled by spherocylinders.  The electrical
connections are implemented as tunneling processes, leading to a more detailed and realistic
description than a discontinuous percolation approach.
We find that the spheres enhance the tunneling conductivity for a
given concentration of rods and that the enhancement increases with rod concentration
into the regime where the conducting network is well established.
By reformulating the network of rods using a critical path analysis, we quantify the effect
of depletion-induced attraction between the rods due to the spheres.
Furthermore, we show that our conductivity data are quantitatively
reproduced by an effective medium approximation, which explicitly relates the
system tunneling conductance to the structure of the rod--sphere fluid.
\end{abstract}

%\pacs{73.40.Gk, 82.70.Dd, 61.20.Ja, 64.60.ah}

\maketitle

\section{Introduction}
\label{intro}

Electrical conductivity in composite materials can be achieved by embedding 
a network of conducting filler particles in an insulating polymeric matrix.
It is well known that the conductivity of such dispersions can be
greatly enhanced
by using filler particles with a large degree of shape anisotropy,
due to the increased excluded volume effects associated with highly non-spherical 
particles \cite{Balberg1984,Bug1985}.
This enhancement in turn makes it possible to reduce significantly
the loading of conductor necessary to establish an adequate level of
electrical connectedness for a given purpose
\cite{Crossman1985, Leung1991, Chatterjee2000, Wang2003, Kyrylyuk2008,Schilling2010}. 

A low filler concentration is often desirable in order to preserve the optical and mechanical
properties of the host insulating medium. Therefore, the rationale behind a large amount of 
experimental work done in the last decade, with particular emphasis on carbon nanotube 
(CNT) fillers dispersed in polymeric matrices, has been to attain high conductivities 
with low filler concentration (see, for example, Refs.~\cite{Bauhofer2009,Dresselhaus2011} for 
comprehensive reviews). 
With this goal in mind, elongated fillers with diameters on the micron scale, such as stainless steel
fibers or metallized glass fibers were already being employed in the early 1980s \cite{Crossman1985}.
In nanocomposites, in addition to CNTs, insulating polymers have also been loaded with short
conducting carbon fibers \cite{Saleh2009} and metal nanowires \cite{Gelves2006, Weber1997}.

Recently it has been reported that the conductivity of a CNT-based nanocomposite can be
greatly enhanced at a given concentration of CNTs by adding small quantities of 
conductive latex depletants, which lead to attractive depletion interactions between the 
CNTs \cite{Kyrylyuk2011}.  Weak attractive 
interactions induced by surfactant micelles have also been shown to strongly enhance the 
dielectric constant in dispersions of CNTs in aqueous solution \cite{Vigolo2005}.

On the theoretical side, by modeling fibrous fillers as rigid cylinders, capped cylinders 
or spheroids, conductivity has usually been interpreted in terms of percolation theory
using both analytical \cite{Balberg1984,Bug1985,Leung1991,Chatterjee2008,Otten2009} and numerical
\cite{Balberg1984-87,Neda1999,Foygel2005,Berhan2007} methods. In the percolation
approach, the filler particles are considered as being coated with a penetrable contact 
shell, and
connectivity between two particles is established at the point where their
shells overlap. The system becomes electrically conducting at the
percolation threshold $\phi_c$, which is the critical volume fraction of filler particles
above which a system-spanning cluster of interconnected fillers is always found.
The calculated values of $\phi_c$ are thus used to estimate the conductor--insulator
transition of real composites.

The theoretical studies mentioned above deal with dispersions of penetrable or impenetrable 
pure rod-like fluids without taking into account further sources of interaction. 
In the framework of percolation theory, the effects of depletion interactions have been 
considered in Ref.~\cite{Kyrylyuk2008}, where particle interactions are mediated by a square well 
attractive potential and in Refs.~\cite{Wang2003, Schilling2007, Kyrylyuk2011}, where hard 
spherical depletant particles are added explicitly into the insulating phase.
This work has revealed an interplay of opposing effects.  On the
one hand, depletion encourages attraction between the surfaces of the rods, which
leads to a higher density of overlapping shells and a lower percolation density.  On the
other hand, depletion also enhances mutual alignment of neighboring rods, decreasing the
spatial extent of the clusters and tending to raise the percolation threshold.  However,
the former effect has been shown to be the stronger \cite{Schilling2007} and the net
result of depletion is to lower the density required for percolation.

Theoretical studies on the conductivity of networks of attractive rods have so far
concentrated on determining the location of the percolation threshold for a particular
choice of contact shell thickness.  For a real composite, 
however, the details of charge transport across the network at filler concentrations 
above the conductor-insulator transition are crucial.
Motivated by the recent reinterpretation of tunneling transport for composites loaded with
anisotropic particles in terms of a global tunneling network \cite{Ambrosetti2010a}, we 
present here a numerical analysis of the conductivity of mixtures composed of
conducting spherocylinders and insulating spheres.  We show that the addition of a low
concentration of spherical depletants
enhances the tunneling conductivity of the sub-system of conducting rods with respect 
to the case without depletants.  The magnitude of this effect rises as the
concentration of rods is increased and continues to enhance the conductivity in the
regime where the conducting network is well established, eventually falling off
slightly at the highest concentrations included in this work.
By computing the critical distance $\delta_c$ associated with
the critical path approximation for system conductivity, we observe that the average
of the relevant interparticle distance is lowered by depletion,
thereby favoring electronic transport. 

We also show that the numerical conductivity results can be reproduced by applying an
effective medium approximation, previously formulated for the case of spherical fillers, 
here generalized to the case of fluids of conducting spherocylinders.  We find that the 
behavior of the effective conductance and the effects of the depletant particles can be 
fully understood in terms of an appropriate pair distribution function dependent on the 
distance between the cores of two spherocylinders. 

\begin{figure}[t]
\begin{center}
\includegraphics[scale=0.24,clip=true]{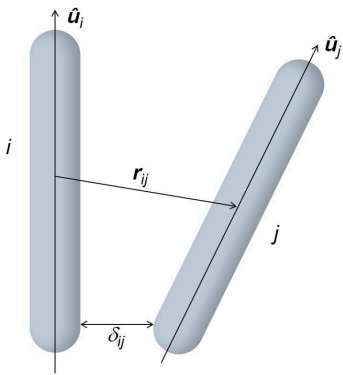}
\caption{(Color online) Schematic representation of two spherocylinders. The position vector $\mathbf{r}_{ij}$
identifies the position of the center of mass of rod $j$ with respect to the center of mass of rod $i$.
$\mathbf{\hat{u}}_i$ and  $\mathbf{\hat{u}}_j$ are unit vectors along the axes of rods
$i$ and $j$, respectively.
$\delta_{ij}$ is the shortest distance between the surfaces of the two rods.}\label{fig1}
\end{center}
\end{figure}

\section{Model and simulations}
\label{sec:model}

We model mixtures of conducting rod-like fillers and insulating spherical depletants by a binary
system composed of $N_r$ impenetrable spherocylinders of length $L$ and diameter $D$, and $N_s$
hard spheres. Since in experiments the depletant particles have diameters comparable to those of
CNTs \cite{Vigolo2005}, we set the sphere diameter equal to $D$. We also require that the
two species of particles are mutually impenetrable.
By representing the spherical depletants explicitly, rather than
implicitly through an effective potential, we ensure that all many-body effects are, by
construction, included.
The volume fractions for spherocylinders and spheres dispersed in a cubic box of edge
$\mathcal{L}$ are, respectively, $\phi_r=\rho_r v_r$ and $\phi_s=\rho_s v_s$.  Here,
$v_r=(\pi/6) D^3+(\pi/4)LD^2$ and $v_s=(\pi/6) D^3$ are the volumes of one spherocylinder and
one sphere, respectively, while $\rho_r=N_r/\mathcal{L}^3$ and $\rho_s=N_s/\mathcal{L}^3$ are
the corresponding number densities.  
We simulate systems of $1000\leq N_r\leq 2888$ sphereocylinders with $L/D=5$ ($\mathcal{L}\geq 3L$)
and $1000\leq N_r\leq 2780$ spherocylinders with $L/D=10$ ($\mathcal{L}\geq 4L$).
Depending on $\phi_s$, the number of spheres ranges between $N_s=8500$ and
$N_s=3.2\times 10^5$.

To generate equilibrium dispersions of the mixture, we start by introducing the rods and
the spheres into the periodic simulation cell by random sequential addition, 
i.e., simply assigning a uniformly distributed random
position and orientation to each particle in turn and
accepting any insertion that does not lead to an overlap with
particles that have been already placed. We test the overlap between two spherocylinders $i$ and $j$
by computing the shortest distance $d_{ij}$ between the two line segments coinciding with
the axes of the spherocylinders \cite{Allen1993}. If $d_{ij}$ is
less than $D$, the two spherocylinders overlap.
The shortest distance is computed by minimizing $\vert \mathbf{r}_{ij}+\lambda_i \mathbf{\hat{u}}_i
-\lambda_j \mathbf{\hat{u}}_j\vert^2$ with respect to $\lambda_i$ and $\lambda_j$
in the range $-L/2$ to $L/2$, where
$\mathbf{r}_{ij}$ is the displacement
vector between the two rod centers, and $\mathbf{\hat{u}}_i$
and $\mathbf{\hat{u}}_j$ are unit vectors along the axes of rods $i$ and $j$, respectively 
(see Fig.~\ref{fig1}).
We then perform a standard Metropolis Monte Carlo equilibration consisting of
trial translational and rotational moves for each rod and sphere.  Given the simple hard core
potential, trial moves that do not lead to hard-core overlap of the particles are always
accepted, while moves that do generate overlaps are rejected.  We monitor any global alignment
of the rods by means of the nematic order parameter $S$ which is the largest eigenvalue of the
tensor $\mathbf{Q}=(2N_r)^{-1}\sum_{i}^{N_r}(3 \mathbf{\hat{u}}_i\mathbf{\hat{u}}_i-\mathbf{1})$, 
where $\mathbf{1}$ is the identity matrix \cite{deGennes1995}.  For the range of
concentrations of rods and spheres used in this study, $S$ is always close to zero,
which indicates that the dispersions of spherocylinders are isotropic. 

In modeling the electron transfer between spherocylinders, we neglect charging and Coulomb
interaction effects and assume that the conductance $g_{ij}$ between any two spherocylinders
$i$ and $j$ is given by single-electron tunneling processes:
\begin{equation}
\label{eq:dep_gij}
g({\delta_{ij}})=g_0 \exp \left(-\frac{2\delta_{ij}}{\xi}\right).
\end{equation}
Here, $\delta_{ij}$ is a shorthand notation for
$\delta_{ij}=\delta_{ij}(\mathbf{r}_{ij};\mathbf{\hat{u}}_i,\mathbf{\hat{u}}_j)$ which is 
the minimal distance between the surfaces of rods $i$ and $j$, given their
relative position and orientation vectors, as schematically illustrated in Fig.~\ref{fig1}.  
The tunneling decay length $\xi$ depends on the electronic potential barrier between two 
spherocylinders and its value ranges from a fraction of a nanometer to a few nanometers.  
In Eq.~\eqref{eq:dep_gij} we set the conductance prefactor $g_0$ equal to unity, so that 
the conductance between two touching spherocylinders is $g(0)=1$.  

Equation~\eqref{eq:dep_gij} captures the dominant dependence of the conductance between two
rods on their relative positions and orientations, i.e., the exponential decay as a
function of the
single distance of closest approach $\delta_{ij}$ of the rods.  However, this simple
formula neglects any explicit dependence of the conductivity on the mutual alignment of
the rods.  The relative orientation does not affect the exponential dependence on $\delta_{ij}$, but
strong alignment can enhance the pre-exponential factor.  This enhancement becomes
significant only when the centers of the two rods are close and
the angle between the two rods is small.  A detailed analysis of the pairwise conductivity
of rods will be the subject of a separate presentation \cite{rodtunnel}.

\begin{figure}[t]
\begin{center}
\includegraphics[scale=1.0,clip=true]{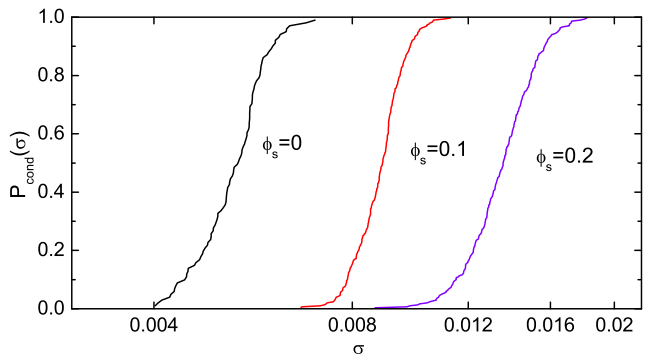}
\caption{(Color online) The cumulative conductivity distribution $P_{\rm cond}(\sigma)$ for systems of
conducting spherocylinders of aspect-ratio $L/D=5$ and volume fraction $\phi_r=0.135$ mixed
with different concentrations $\phi_s$ of insulating spherical particles. 
The tunneling decay length is fixed at $\xi/D=0.2$.}\label{fig2}
\end{center}
\end{figure}

\section{Network conductivity}
\label{sec:sd_cond}

To obtain the overall conductivity $\sigma$, we construct for each 
realization of the system a conductance network by assigning the tunneling conductances 
from Eq.~\eqref{eq:dep_gij} to each pair of spherocylinders.  The tunneling network does not 
include the spheres, which are insulating.  The effect of the spheres on conductivity is
indirect, through the depletion forces that they induce between the spherocylinders.

\begin{figure*}[t]
\begin{center}
\includegraphics[scale=0.9,clip=true]{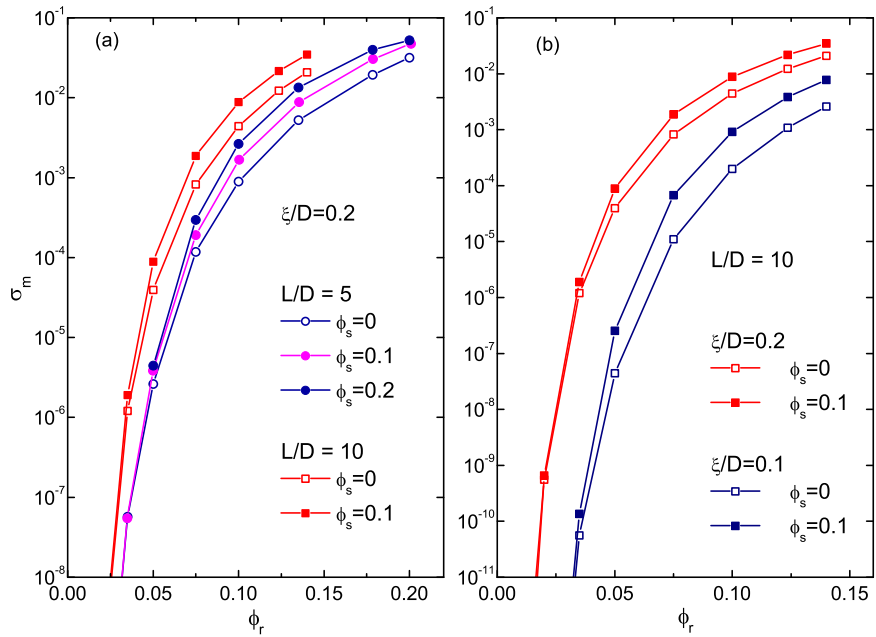}
\caption{(Color online) Conductivity $\sigma_m$ computed from the numerical solution
of the tunneling network equations as a function of the volume fraction $\phi_r$ of
conducting spherocylinders and for different concentrations $\phi_s$ of insulating spheres. 
Results for (a) $\xi/D=0.2$ and $L/D=5$, $10$ and (b) $L/D=10$ and $\xi/D=0.1$, $0.2$.}\label{fig3}
\end{center}
\end{figure*}

Despite involving only the spherocylinders, the network is composed of $N_r (N_r-1)/2$ tunneling
bonds, which renders the numerical solution of the overall conductivity computationally demanding
for the system sizes considered here.  We therefore reduce the number of bonds, 
and so the coordination number of each node, by eliminating from the
network the tunneling conductances associated with pairs of spherocylinders that are sufficiently
far apart not to contribute to the overall conductivity \cite{Nigro2011,Nigro2012}. 

We solve the Kirchhoff equations of the reduced network by combining exact numerical decimation with
a preconditioned conjugate gradient algorithm \cite{Fogelholm1980,Batrouni1988,Golub1996,Johner2009}.
The decimation algorithm uses exact transformations to eliminate nodes from the 
network and to update the conductances adjacent to the eliminated nodes \cite{Fogelholm1980}, as discussed
in the Appendix.  We iteratively decimate
the network starting from the nodes with the lowest coordination number until a single
conductance is left, whose value coincides with the conductance of the original network.
For configurations such that the computational time for the node decimation is too large, we switch to the 
conjugate gradient method with Cholesky preconditioning (see, e.g., Refs.~\cite{Batrouni1988,Golub1996}) 
applied to the partially decimated network. 
We have applied this procedure to $300$ realizations of each system.
For each equilibrium configuration we obtain the
conductivity from $\sigma=GD/\mathcal{L}$, where $G$ is the conductance of the reduced network
for a unit voltage drop applied to two opposite faces of the simulation box.  From the sample
of configurations at a given combination of $\phi_r$, $\phi_s$, $\xi/D$, and $L/D$, we construct
the cumulative conductivity distribution function $P_{\rm cond}(\sigma)$, 
which gives the probability of finding conductivities less than or equal to 
$\sigma$ over all realizations. This function is
shown in Fig.~\ref{fig2} for three concentrations of depletants with rods of aspect ratio
$L/D=5$, packing fraction $\phi_r=0.135$, and tunneling length $\xi/D=0.2$.

To compare conductivities at different concentrations of spheres or rods, we concentrate 
on the conductivity $\sigma_m$ that marks the mid-point of the cumulative distribution,
$P_{\rm cond}(\sigma_m)=1/2$.  $\sigma_m$ increases with the concentration of
depletant spheres, as seen by comparing the results for $\phi_s=0$, $0.1$, and $0.2$ shown
in Fig.~\ref{fig2}.  This trend is confirmed in Fig.~\ref{fig3}(a), where we show $\sigma_m$
as a function of the rod volume fraction $\phi_r$ for $L/D=5$ 
and $L/D=10$ and for different depletant concentrations $\phi_s$.
For a given $\phi_s$, the enhancement of $\sigma_m$
is negligible at low rod concentrations $\phi_r$, where small changes in the fluid
structure have little effect on network connectivity, and increases
with rod concentration as the network emerges. In this regime, the increase is such that
systems with $L/D=5$ and $\phi_s=0.2$ have conductivities
approaching those of hard rods with $L/D=10$ and no depletants.
There is a slight drop in the level of enhancement at the very largest rod concentrations, 
since even a network of pure rods is highly ramified at such density. 
For given densities of both rods and spheres, the depletion-induced increase of $\sigma_m$
becomes stronger as the tunneling decay length decreases, as seen by comparing the results for
$\xi/D=0.2$ and $0.1$ shown in Fig.~\ref{fig3}(b).

We note that in our calculations we neglect any contribution stemming
from intrinsic conductivity $\sigma_{\rm ins}$ of the insulating matrix, which in real composites is small but finite
at nonzero temperatures. The inclusion of $\sigma_{\rm ins}$ would prevent the conductivity of the
composite system from dropping to zero as $\phi_r\rightarrow 0$ by limiting its value at $\sigma_{\rm ins}$. 
In this respect, the conductor-insulator transition point can be estimated
by the value of $\phi_r$ such that $\sigma_m\approx\sigma_{\rm ins}$ \cite{Ambrosetti2010a,Nigro2012}.

\begin{figure}[b]
\begin{center}
\includegraphics[scale=1.0,clip=true]{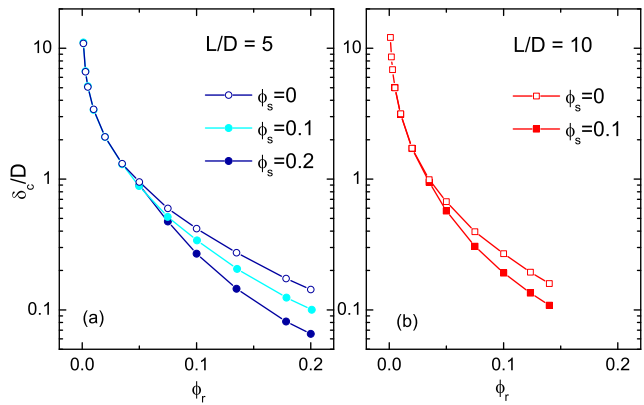}
\caption{(Color online) Critical distance $\delta_c$ as a function of the volume fraction $\phi_r$ of
spherocylinders with (a) $L/D=5$ and (b) $L/D=10$, and for different volume fractions $\phi_s$ 
of the depletant spheres. }\label{fig4}
\end{center}
\end{figure}

To explore the mechanism causing the conductivity enhancement highlighted in Fig.~\ref{fig3},
we calculate the critical distance $\delta_c$, which is defined as the smallest distance
such that the subnetwork of bonds satisfying $\delta_{ij}\leq \delta_c$ still spans the entire sample.
The interest in evaluating $\delta_c$ stems from the observation of Ref.~\cite{CPApapers} that when the values of the single
conductances $g(\delta_{ij})$ vary over many orders of magnitude, and so when
the $\delta_{ij}$ distances span several multiples of the tunneling decay length $\xi$,
the network conductance is dominated by the characteristic bond conductance 
$g(\delta_c)=g_0\exp(-2\delta_c/\xi)$. In this situation, the network conductivity 
is well estimated by the critical path approximation (CPA) formula \cite{CPApapers}:
\begin{equation}
\label{CPA}
\sigma_\textrm{cpa}=\sigma_0\exp\!\left(-\frac{2\delta_c}{\xi}\right),
\end{equation}
where $\sigma_0$ is a slowly varying function of $\delta_c$ and can be considered constant.  
Using this expression, we can represent the conductivity as
a function of $\delta_c$, thereby making a connection with approaches based on percolation theory.
However, unlike the usual percolation approach of imposing a fixed contact shell thickness,
the CPA uses a variable thickness whose value responds to the structure of the fluid.
It follows that Eq.~\eqref{CPA} predicts through $\delta_c$ an implicit dependence of the
conductivity on the parameters of the network of conducting particles, in contrast to the
usual percolation approach in which the information on the fluid structure is contained
only in the percolation threshold $\phi_c$.

We calculate $\delta_c$ following the method described in
Refs.~\cite{Nigro2011,Nigro2012,Nigro2013a,Nigro2013b}, which consists of constructing the percolation
probability $P_{\rm perc}(\delta)$ from the values of the percolation distance
$\delta$ calculated for all realizations of the system with a given combination of
rod and sphere concentrations.
From the condition $P_{\rm perc}(\delta_c)=1/2$, which provides a robust estimate of the critical
distance \cite{Nigro2012}, we find that the effect of the depletant particles is to lower $\delta_c$
compared to the case without depletants ($\phi_s=0$), as shown in Figs.~\ref{fig4}(a) and \ref{fig4}(b).  
Furthermore, this lowering becomes asymptotically negligible as the concentration
of spherocylinders goes to zero. 

The behavior of $\delta_c$ shown in Fig.~\ref{fig4} can be understood in terms of the hard-core
repulsion between rods and spheres.  When two spherocylinders approach closely, the volume available
to spheres increases, inducing an effective attraction between the spherocylinders.  In this way,
percolation of the system of rods is established for shorter distances compared to the 
case without spheres, thus reducing $\delta_c$. 

\begin{figure}[t]
\begin{center}
\includegraphics[scale=1.0,clip=true]{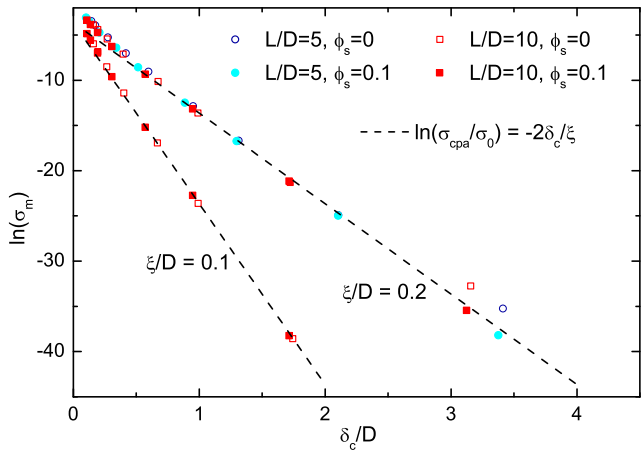}
\caption{(Color online) Conductivity values of Fig.~\ref{fig3} replotted as a function of
$\delta_c/D$, where $\delta_c$ is the calculated critical distance.  The dashed lines are linear
fits to Eq.~\eqref{CPA}.  For all cases the prefactor $\sigma_0$ is $0.026$.}\label{fig5}
\end{center}
\end{figure}

By combining the results in Fig.~\ref{fig4} with the CPA formula given in Eq.~\eqref{CPA},
we see that the lowering of $\delta_c$ induced by the depletant spheres directly translates
into a corresponding enhancement of $\sigma_\textrm{cpa}$.  By identifying
$\sigma_\textrm{cpa}$ with $\sigma_m$, the depletion interaction effect thus explains
the enhancement of the sample conductivity shown in Fig.~\ref{fig3}.  To test the accuracy
of the CPA conductivity, we plot in Fig.~\ref{fig5} the calculated $\ln(\sigma_m)$ values
of Fig.~\ref{fig3} as a function of $\delta_c/D$ (symbols) and compare them
with Eq.~\eqref{CPA} using $\sigma_0=0.026$ (dashed lines), obtained from a fit at low
conductivity.  The exponential behavior of $\sigma_\textrm{cpa}$ reproduces 
quantitatively the functional dependence of $\sigma_m$ for $\delta_c/D\gtrsim 0.5$, while the CPA
becomes less accurate as $\delta_c/D\rightarrow 0$.  In general, the CPA is not expected
to reproduce correctly the tunneling conductivity when $\delta_c\lesssim \xi$ since in this case
the $\delta_{ij}$ distances would not be widely distributed with
respect to $\xi$, as previously discussed.  Although $\sigma_m$ deviates from a simple exponential function
of $\delta_c$ at low $\delta_c$, data for different aspect ratios and depletant concentrations
still collapse onto the same curve for a given tunneling length $\xi$, as shown by the
two sets of data for different $\xi/D$ in Fig.~\ref{fig5}. 

\section{Effective-medium approximation}
\label{sec:sd_EMA}

An alternative approach to the CPA to describe tunneling transport in suspensions of conducting
particles is provided by the effective medium approximation (EMA) applied to the network of tunneling
conductances of Eq.~\eqref{eq:dep_gij}.  In contrast to the CPA, the EMA approach does not rely on the
definition of a percolation quantity such as $\delta_c$, and applies also to systems with typical
inter-particle distances lower than the tunneling decay length $\xi$.  Furthermore, within the two-site 
approximation, the EMA explicitly relates the system conductance to the pair distribution function of the
conducting particles, thus emphasizing the role of the fluid structure in tunneling transport.

We apply the EMA to the fluid of conducting rods and spherical depletants following the method
described in Ref.~\cite{Grimaldi2011}, in which each pairwise conductance $g(\delta_{ij})$ in the
network of $N_r$ spherocylinders is replaced by an effective conductance $\bar{g}$ that is
independent of $\delta_{ij}$.  Requiring equivalence between the average resistance of the 
original tunneling network and that of the effective one, the conductance 
$g^*=N_r \bar{g}/2$ between any two nodes of the network is
calculated from the solution of the following equation:
\begin{equation}
\label{EMA1}
\frac{1}{N_r}\left\langle\sum_{i=1}^{N_r}\sum_{j\neq i}^{N_r}\frac{g(\delta_{ij})}{g^*+g(\delta_{ij})}\right\rangle=2,
\end{equation}
where the angular brackets indicate an ensemble average over configurations.  
The two-point EMA conductance $g^*$ is independent of the network size, and can thus be regarded as
a measure of the system conductivity. This size independence is
a consequence of the completeness of the EMA network, in which
all nodes are connected with identical resistances \cite{Wu2004}.

Since the tunneling conductances depend
only on the distances $\delta_{ij}$ between the spherocylinders, we recast
Eq.~(\ref{EMA1}) in a form involving the probability function of the distance
between pairs of rods, $P_\textrm{pair}(\delta)$, which is readily obtained by binning
of $\delta_{ij}$ during the simulations:
\begin{equation}
\label{EMA2}
\rho_r\int_0^\infty\! d\delta 
\frac{P_\textrm{pair}(\delta)}{(g^*/g_0)\exp(2\delta/\xi)+1}=2.
\end{equation} 

\begin{figure}[t]
\begin{center}
\includegraphics[scale=1.0,clip=true]{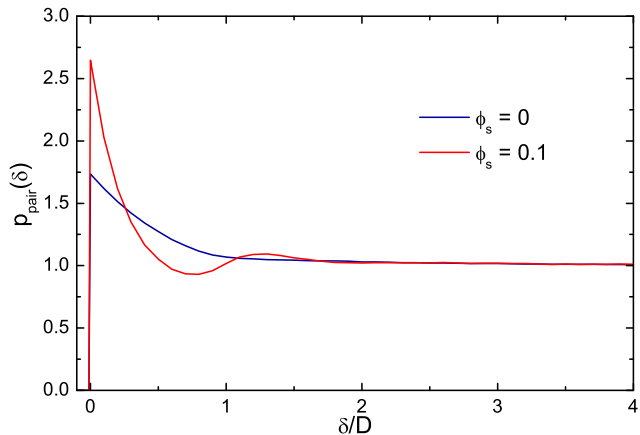}
\caption{(Color online) The distribution function of distance $\delta$
between the surfaces of pairs of spherocylinders relative to the ideal case,
$p_\textrm{pair}(\delta)$, for aspect ratio $L/D=10$ at $\phi_r=0.14$ as a function
of $\delta/D$.}\label{fig6}
\end{center}
\end{figure}

The function $P_{\rm pair}(\delta)$ in our mixture of hard-core particles is best
visualized by normalizing it with respect to the corresponding function $P_{\rm id}(\delta)$ of
ideal (fully penetrable) spherocylinders at the same density.  The resulting quantity, 
$p_{\rm pair}(\delta)=P_{\rm pair}(\delta)/P_{\rm id}(\delta)$, coincides with the average correlation
function introduced in Ref.~\cite{Boublik1976}. The ideal distribution can be
derived from
\begin{displaymath}
P_{\rm id}(\delta) = \rho_r S_{\rm exc}(\delta),
\end{displaymath}
where $S_{\rm exc}(\delta)$ is the surface area of the orientationally averaged excluded
volume $V_{\rm exc}(\delta)$ for two rods with cores constrained to a fixed closest distance
$\delta$ \cite{Boublik1976,Vega1989}.  Hence,
$S_{\rm exc}(\delta) = dV_{\rm exc}(\delta)/d\delta$, where \cite{Balberg1984}
\begin{displaymath}
V_\textrm{exc}(\delta)=\frac{4\pi}{3}(D+\delta)^3+2\pi(D+\delta)^2L+\frac{\pi}{2}(D+\delta)L^2.
\end{displaymath}

In Fig.~\ref{fig6} we show the numerical $p_{\rm pair}(\delta)=P_{\rm pair}(\delta)/P_{\rm id}(\delta)$
averaged over $300$ realizations for $\phi_r=0.14$, $L/D=10$
with $\phi_s=0$ and $0.1$.  
The effect of the depletants is manifest in the increased
oscillations of $p_\textrm{pair}(\delta)$ \cite{Schilling2007} for the case $\phi_s=0.1$, 
indicating a strong spatial correlation due to the reduction of the interparticle distances.

\begin{figure}[t]
\begin{center}
\includegraphics[scale=1.0,clip=true]{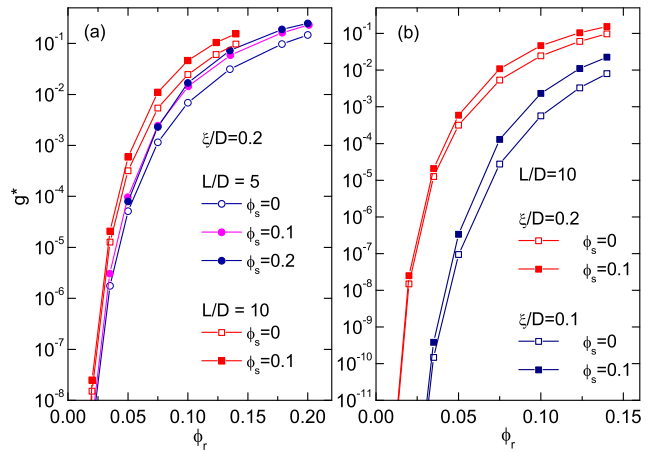}
\caption{(Color online) Two-point EMA conductance $g^*$ obtained from the numerical solution of 
Eq.~\eqref{EMA2} as a function of $\phi_r$ and for different volume 
fractions $\phi_s$ of insulating spheres. results for
(a) $\xi/D=0.2$ and $L/D=5$, $10$, and (b) $L/D=10$ and $\xi/D=0.1$, $0.2$.}\label{fig7}
\end{center}
\end{figure}

Using the measured $P_{\rm pair}(\delta)$ to solve Eq.~\eqref{EMA2} numerically for $g^*$, we obtain
the plots in Fig.~\ref{fig7}.  The increased population of rods at short distances, induced by
the depletant spheres and highlighted in Fig.~\ref{fig6}, promotes
tunneling processes at short distances, which result in larger EMA conductances $g^*$
compared to the depletant-free case.
Figure~\ref{fig7} indeed shows $g^*$ values systematically enhanced by the introduction of
depletants for fixed $\phi_r$ and $L/D$.
Furthermore, the EMA results in Fig.~\ref{fig7} reproduce accurately the conductivity 
behavior obtained by the full numerical solution of the tunneling network 
shown in Fig.~\ref{fig3}. 

\section{Conclusions}
\label{concl}

Our simulation results show that the tunneling conductivity of fluids of rods is systematically enhanced
by the addition of insulating spheres, even in the regime where the conducting network is well
established.  The mechanism of transport enhancement is the depletion effect of the spheres,
which induces an effective attraction between the conducting rods.

Previous work \cite{Kyrylyuk2008,Wang2003,Kyrylyuk2011,Schilling2007}
has already established that the percolation threshold with an arbitrarily
chosen contact shell thickness is lowered by depletion attraction.  However,
we have taken the analysis further in the present work in two ways.
First, within the framework of percolation
theory, the critical path approximation not only shows and quantifies how
the minimum distance required for
a sub-system of rods to span the system decreases with increasing depletant concentration,
but also provides a good approximation to the conductivity over a wide range of parameters,
thus clarifying the role of depletants through the reduction of the dominant tunneling distances.

Second, moving on from approaches based on percolation theory,
we were also able to reproduce the simulation-based conductivity results by applying a simple
effective medium approximation to the tunneling network formed by the rods.  Using the pair
distribution function for the distance of closest approach of the spherocylinders, computed
from our simulations, we found that the resulting effective conductance has the same functional
dependence on the system parameters as was observed from the full calculation of the conductivity.

To conclude, let us comment on a possible extension of our work.
As pointed out in Sec.~\ref{sec:model}, the model of electron transfer of Eq.~\eqref{eq:dep_gij} neglects
possible effects of mutual rod alignment on the probability of tunneling between rods. 
Although this approximation is valid for isotropic dispersions of rods,
such as the ones considered in the present work, for systems with a
high degree of rod alignment the dependence of tunneling on the relative rod orientation should be taken 
into consideration \cite{rodtunnel}. In particular, the tunneling conductance is expected to be larger
when the rods are aligned, and so it may compete with the reduction of connectivity when the concentration
of depletants is sufficiently large to induce nematic order. Further studies in this direction are
computationally demanding, but would shed light on still unexplored issues.

B. N. acknowledges support by the Swiss National Science Foundation (Grant No. 200020-135491).

\appendix
\section{Algorithm for decimation of conductance networks}

\begin{figure}[t]
\begin{center}
\includegraphics[scale=0.33,clip=true]{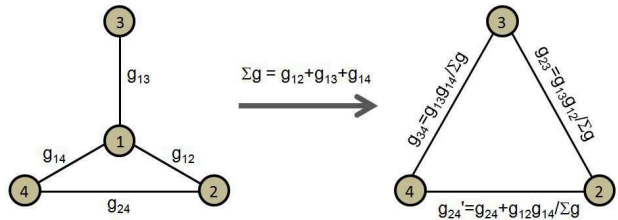}
\caption{Schematic illustration adapted from Ref.~\cite{Johner2009} of the star-mesh 
transformation applied to a network
with three terminal nodes. The transformation acts by (i) removing the internal node $1$ and the conductances
between the nodes $2$, $3$, and $4$, and (ii) inserting new conductances between the pairs $(2,3)$, $(2,4)$, 
and $(3,4)$. The new conductance between $2$ and $4$ is added in parallel to the conductance $g_{24}$ of the 
original network.}\label{fig8}
\end{center}
\end{figure}

The decimation algorithm \cite{Fogelholm1980} is based on the star-mesh transformation, well known 
in electrical circuit theory,
which eliminates one node from the network at the price of inserting new conductances between certain remaining
nodes. The new conductances are chosen as to ensure electrical equivalence between the modified and the 
original networks. If the node $i$ is directly connected to $n$ other nodes through conductances $g_{il}$, 
the transformation eliminates the node $i$ and inserts $n(n-1)/2$ new conductances between each pair of the nodes 
which were originally connected to $i$. If one such pair
has node indices $j$ and $k$, the new conductance is
\begin{equation}
\label{starmesh}
g_{jk}=\frac{g_{ij}g_{ik}}{\sum_{l} g_{il}},
\end{equation}
where the sum runs over all neighbors $l$ of node $i$. If the original network had pairs of neighbors already
connected, the new conductances are inserted in parallel to the existing ones. An illustration of the
star-mesh transformation for $n=3$ is shown in Fig.~\ref{fig8}.

We decimate the network by sequentially applying the star-mesh transformation to all nodes except those
representing the electrodes. Ideally, by iterative decimation,
the whole network is replaced by just one
conductance between the electrodes, which is the equivalent conductance of the network. 
However, for networks with high degrees of coordination per node, as in systems with large concentrations of rods,
the decimation procedure increases enormously the number of added conductances as the nodes are 
eliminated, considerably increasing the computational time. In this case, we implement the conjugate 
gradient method with Cholesky preconditioning \cite{Batrouni1988,Golub1996}, which performs better
when applied to resistor networks with large coordination numbers.

\end{document}